\documentstyle[11pt,adassconf]{article}  % Leave intact

\begin{document}   % Leave intact

%-----------------------------------------------------------------------
%			    Paper ID Code
%-----------------------------------------------------------------------
% Enter the proper paper identification code.  The ID code for your
% paper is the session number associated with your presentation as
% published in the official ADASS 2000 conference program.  You can
% find this number locating your abstract in the printed program
% that you received at the meeting or on-line via
% http://cadcwww.hia.nrc.ca/adass_2001; the ID code is the letter-number
% sequence proceeding the title of your presentation.
%
% This will not appear in your paper; however, it allows different
% papers in the proceedings to cross-reference each other.
%
% EXAMPLE: \paperID{O-3}
% EXAMPLE: \paperID{P-56}
%
% Note that you should only have one \paperID, and it should not
% include a trailing period.  

\paperID{O-4}

%-----------------------------------------------------------------------
%		            Paper Title 
%-----------------------------------------------------------------------
% Enter the title of the paper.
%
% EXAMPLE: \title{A Breakthrough in Astronomical Software Development}
% 

\title{Theory in a Virtual Observatory}

%-----------------------------------------------------------------------
%		          Authors of Paper
%-----------------------------------------------------------------------
% Enter the authors followed by their affiliations.  The \author and
% \affil commands may appear multiple times as necessary (see example
% below).  List each author by giving the first name or initials first
% followed by the last name.  Authors with the same affiliations
% should grouped together. 
%
% EXAMPLE: \author{Raymond Plante, Doug Roberts, 
%                  R.\ M.\ Crutcher\altaffilmark{1}}
%          \affil{National Center for Supercomputing Applications, 
%                 University of Illinois Urbana-Champaign, Urbana, IL
%                 61801}
%          \author{Tom Troland}
%          \affil{University of Kentucky}
%
%          \altaffiltext{1}{Astronomy Department, UIUC}
%
% In this example, the first three authors, "Plante", "Roberts", and
% "Crutcher" are affiliated with "NCSA".  "Crutcher" has an alternate 
% affiliation with the "Astronomy Department".  The fourth author,
% "Troland", is affiliated with "University of Kentucky"

\author{Peter Teuben}
\affil{Astronomy Department, University of Maryland, College Park, MD}

\author{Dave DeYoung}
\affil{NOAO, Tucson, AZ}

\author{Piet Hut}
\affil{Institute for Advanced Study, Princeton, NJ}
% 08540}

\author{Stuart Levy}
\affil{National Center for Supercomputing Applications, 
        University of Illinois Urbana-Champaign, Urbana, IL}

\author{Jun Makino}
\affil{Department of Astronomy,
  The University of Tokyo,
  Bunkyo-ku,
  Tokyo 113-0033,
  JAPAN}
\author{Steve McMillan}
%\affil{   Drexel University, Philadelphia, PA 19104}
\affil{ Department of Physics and Atmospheric Science,
         Drexel University, Philadelphia, PA}
\author{Simon Portegies Zwart\altaffilmark{1}}
\affil{Massachusetts Institute of Technology, Cambridge, MA}

\author{Shawn Slavin}
\affil{Department of Astronomy, Indiana University, Bloomington, IN}

\altaffiltext{1}{Hubble Fellow}

%-----------------------------------------------------------------------
%			 Contact Information
%-----------------------------------------------------------------------
% This information will not appear in the paper but will be used by
% the editors in case you need to be contacted concerning your
% submission.  Enter your name as the contact along with your email
% address.
% 
% EXAMPLE:  \contact{Dennis Crabtree}
%           \email{crabtree@cfht.hawaii.edu}
%

\contact{Peter Teuben}
\email{teuben@astro.umd.edu}

%-----------------------------------------------------------------------
%		      Author Index Specification
%-----------------------------------------------------------------------
% Specify how each author name should appear in the author index.  The 
% \paindex{ } should be used to indicate the primary author, and the
% \aindex for all other co-authors.  You MUST use the following
% syntax: 
%
% SYNTAX:  \aindex{LASTNAME, F. M.}
% 
% where F is the first initial and M is the second initial (if
% used).  This guarantees that authors that appear in multiple papers
% will appear only once in the author index.  
%
% EXAMPLE: \paindex{Crabtree, D.}
%          \aindex{Manset, N.}        
%          \aindex{Veillet, C.}        

\paindex{Teuben, P.J.}
\aindex{DeYoung, D.}
\aindex{Hut, P.}
\aindex{Levy, S.}
\aindex{Makino, J.}
\aindex{McMillan, S.}
\aindex{Portegies Zwart, S.}
\aindex{Slavin, S.}

%-----------------------------------------------------------------------
%			Subject Index keywords
%-----------------------------------------------------------------------
% Enter up to 6 keywords describing your paper.  These will NOT be
% printed as part of your paper; however, they will be used to
% generate the subject index for the proceedings.  There is no
% standard list; however, you can consult the indices for past ADASS
% proceedings (http://iraf.noao.edu/ADASS/adass.html). 
%
% EXAMPLE:  \keywords{visualization, astronomy: radio, parallel
%                     computing, AIPS++, Galactic Center}
%
% In this example, the author noticed that "radio astronomy" appeared
% in the ADASS VII Index as "astronomy" being the major keyword and
% "radio" as the minor keyword.


%-----------------------------------------------------------------------
%			       Abstract
%-----------------------------------------------------------------------
% Type abstract in the space below.  Consult the User Guide and Latex
% Information file for a list of supported macros (e.g. for
% typesetting special symbols). 

\begin{abstract}          % Leave intact

During the last couple of years, observers have started to make plans for
a Virtual Observatory, as a federation of existing data bases,
connected through levels of software that enable rapid searches,
correlations, and various forms of data mining.  We propose to extend
the notion of a Virtual Observatory by adding archives of simulations,
together with interactive query and visualization capabilities, as
well as ways to simulate observations of simulations
in order to compare them with observations. For this
purpose, we have already organized two small workshops, earlier in
2001, in Tucson and Aspen.  We have also provided concrete examples
of theory data, designed to be federated with a Virtual Observatory.
These data stem from a project to construct an archive for
our large-scale simulations using the GRAPE-6
(a 32-Teraflops special purpose computer for stellar dynamics).  We
are constructing 
interfaces by which remote observers can observe these
simulations.  In addition, these data will enable detailed comparisons
between different simulations.

\end{abstract}

% Slides in this 10 minute adass talk:
% 1) theory in a VO
% 2) outline
% 3) G6 and baby G6
% 4) VO and theory?
% 5) galaxy modeling
% 6) galaxy modeling
% 7) theory in a VO
% 8) theory in a VO
% 9) toy model (1)
% 10) toy model (2)
% 11) technical challenges
% 12) conclusions

\section{Introduction}

There are numerous types of theoretical data which, if integrated in a
VO, will without doubt enhance its scientific capabilities. 
%In this paper we will limit ourselves mostly to simulation type data.
Although it has been stressed 
the VO itself is not intended to be a remote observatory,
some branches in the theory part of a VO could very well emulate such
behavior.  One can imagine that after an initial selection
from a set of models or a match to an observation, fine-tuning
be done by re-running the models, given enough computer time and
access to software (for an existing example see e.g. Pound et al. 2000).

% Especially as more advanced models become available, ...

It is perhaps instructive to view the theory part of a VO from two
different points of view: that of the theorist and that of the observer.

\section{The Theorist}

What will a theorist find in a VO? He will find a large number of 
models
that can be ``observed''. Observing such models can be done in
several ways. First, one can make
simulated observations of simulation data, and then compare
observations with these models. Given that many models
add the independent time parameter, simulations also add 
the complexity of exploring 4-dimensional histories 
and finding a best match in the time domain. The 3D spatial
information will mostly likely be on a grid, or a discrete
set of points.

A new and largely unused capability of theory data in a VO
will be to compare models with models,
much like observations are compared. This should also result in
improved models, as differences and similarities between
models can quickly be highlighted.

Theorists will also find a variety of 
standard initial conditions or benchmark data in a VO,
which will make it easier to test new algorithms and compare
them to previously generated data.
In addition, one could also argue that besides saving the data,
saving the code that generated the data will be valuable.
Finally, adding theoretical data to a VO will undoubtedly also
spur new data mining and CS techniques.

\section{The Observer}

What will an observer find about theory data in a VO? 
First, models can be selected and compared to
observations, processing those models as though they were
observed with a particular instrument.

Second, theory data can also be used to calibrate observations.
Examples are:
comparing Hipparchos proper motion studies
with a similar analysis applied to simulations, and using stellar evolution
tracks to determine cluster ages from an HRD. 
The added complexity of theoretical data will need new
searching and matching techniques, and thus bring different
type of data mining and computer science to the playing field.

% Theory will also bring different types
% of data mining and CS techniques to the playing field.

\section{Data Collection Toy Model}

In order to develop a better understanding of theory data, we have started
\htmladdnormallinkfoot{collecting}{http://www.manybody.org/} various
types of theory data, mostly simulations in which time is the
independent variable. Some datasets are simple benchmarks, taking
initial conditions for well-known problems in Astrophysics, going back
to the first published benchmark of the IAU 25-body problem (Lecar
1968).

During the IAU 208 conference in Tokyo (Teuben 2002) a survey was 
undertaken amongst practitioners of a well defined subset of theory
data: particle
simulations. These ranged from planetary to cosmological simulations,
and included grid-based as well as particle-based calculations.
One noteworthy find was that a surprisingly large fraction of the theorists
would rather not like to see their data published in a VO, since computers
get faster each year, algorithms get better and data ages 
quickly. Unlike observations, theoretical data often suffer from 
assumptions and thus comparisons can have less meaning than
naively thought.

% ok, this is a break, and as is, suggests that these are the
% conclusions from the survey....

On a technical note, 
simulation data actually do not differ much from observational
data. Most theoretical data sets fall two types: grid based 
(``image'', each datum being the same type) or particle based
(a ``table'' with columns and rows).  
An image can also be seen as a special case of a table.  
In recent years, added
complexities are nested grids, such as in AMR, and the {\tt tdyn}
tables in Starlab's kira code (Portegies Zwart et al. 2000), where
only relevant particles are updated. The {\tt miriad}
uv-data format
is an example where such complexities have also been introduced
to observational data. Defining the header and meta data for
theoretical data will be at least as challenging as that for
observational data.

\section{GRAPE-6 data archive}

The recently completed GRAPE-6 (Hut and Makino 1999, Makino 2002) can
now produce massive datasets with a size of Terabytes for a single run.
In order to handle these data, and to share them with `guest observers',
we have started to set up a data archive (see also the {\tt manybody.org}
web site).  In the near future we plan to start federating our archive
with other theory archives and with the budding Virtual Observatories.

\acknowledgements{We thank the
Alfred P. Sloan Foundation for a grant to Hut for observing
astrophysical computer simulations in the Hayden Planetarium at the
Museum.  We also thank the American Museum of Natural History for
their hospitality.
}


\begin{references}

% grape paper?
\reference Hut, P., and Makino, J. 1999, Science 283, 64

\reference Lecar, M. 1968,
Bull.Astr. 3, 91
in: Colloque sur le problem gravitationnel des N corps

\reference Makino, J. 2002. in
Astrophysical Supercomputing using Particle Simulations, ed.
J. Makino, and P. Hut (San Francisco: ASP), in press

% \reference  Makino, J. 1999.
% in Proc. Numerical Astrophysics, eds. S.M.Miyama, K.Tomisaka, T.Hanawa
% (Kluwer, Dordrect) 240, 407.

% starlab paper
\reference Portegies Zwart, S.F., McMillan, S.L.W., Hut,
         P. \& Makino, J., 2001, MNRAS, 321, 199
% WITS poster
\reference Pound, M.W., Wolfire, M.G., Mundy, L.G., 
Teuben, P.J., \adassix, 628

% NEMO paper
\reference Teuben, P.J. 1994, \adassiv, 398

% partiview poster
\reference Teuben, P.J. et al. 2001, \adassx, 499

% IAU 208 paper
\reference Teuben, P.J. 2002, in
Astrophysical Supercomputing using Particle Simulations, ed.
J. Makino, and P. Hut (San Francisco: ASP), in press


\end{references}
\end{document}